# Splat formation and microstructure of solution precursor thermal sprayed Nb-doped titanium oxide coatings

Daniel Tejero-Martin[1], Zdenek Pala[2], Simon Rushworth[3], Tanvir Hussain[1*]

[1] Faculty of Engineering, University of Nottingham, University Park, Nottingham, NG7 2RD, UK

[2] GE Aviation, Beranovych 65, Prague, Czech Republic

[3] EpiValence Ltd, The Wilton Centre, Redcar, Cleveland, TS10 4RF, UK

*tanvir.hussain@nottingham.ac.uk; +44 115 951 3795

**Keywords**: Solution precursor; HVOF; niobium; titanium oxide; anatase; single splat

**Abstract**

Solution precursor thermal spray can become a breakthrough technology for the deposition of coatings with novel chemistries; however, the understanding of the process that the feedstock material undergoes is still poorly understood when compared to more traditional presentations (i.e. powder and suspension). In this paper, niobium-doped $TiO_2$ coatings were deposited by solution precursor high velocity oxy-fuel spraying, studying its microstructure and phase. It was reported that a lower flame temperature produced a highly porous coating, while the porosity was reduced at higher flame temperature. Investigation of the phase content showed that, contrary to our current understanding, a higher flame power implied an increase of the anatase phase content for solution precursor spray. Three methods were used: Rietveld refinement, peak height and peak area of the x-ray diffraction patterns. Additionally, single splats were analysed, showing that as the precursor travels through the flame, pyrolysis and sintering takes place to form the solid material. These results were used to derive a model of the physico-chemical transformation of the solution precursor. This work proves that solution precursor thermal spray is a promising technique for the deposition of doped ceramic coatings, being the microstructure and phase content controllable through the spraying parameters.

1. Introduction

Titanium oxide coatings are widely used in a number of fields owing to its unique properties, such as a prominent photocatalytic effect (Ref 1), its electrochemical activity (Ref 2), its coherent change in electrical conductivity under gas exposure (Ref 3) or its ability to produce transparent coatings (Ref 4,5). Nevertheless, the properties of titanium oxide (also known as titania) can be tailored to be more beneficial through doping (Ref 6–11). Among those elements, niobium presents some desirable properties that makes it a suitable candidate for the formation of doped titanium oxide coatings. It effectively increases the electrical conductivity which, in addition to the favourable electrochemical properties of $TiO_2$ (such as high capacity and low volume expansion during ion charge/discharge (Ref 12–14)), makes it a viable option as an anode material for high-power Li-ion batteries. The ability to produce transparent coatings has increased its interest regarding its use as a transparent conductive oxide (TCO) coating. In the recent years there has been a growing demand for a replacement of the current TCO industrial reference, $Sn-In_2O_3$, due to limited resources and increasing costs (Ref 15). An increase of the coating conductivity, coupled with high transmittance in the visible range, positions Nb-doped $TiO_2$ as a viable candidate for the next generation of TCOs (Ref 16–18). In addition to the improvement in electrical conductivity, Nb-doped $TiO_2$ coatings have also found application as sensors, as the niobium limits the grain growth and inhibits the anatase to rutile phase transformation to temperatures up to 650 °C (Ref 19), being the rutile phase traditionally considered detrimental for sensing applications (Ref 11,20–23).

Due to the plethora of applications present, production of niobium-doped $TiO_2$ has been reported in the literature using techniques such as sputtering (Ref 18,24,25), pulsed laser deposition (Ref 26,27), atomic layer deposition (Ref 28), spin-coating (Ref 11,29) or aerosol-assisted chemical vapour deposition (Ref 17). At the same time, thermal spraying deposition of titanium oxide has been thoroughly studied before; using powder, suspensions and solution precursors as feedstock material (Ref 30–33). In thermal spray, a heat source is used to melt the feedstock material while a jet carries the molten particles towards the piece to be coated (Ref 34). In the case of high velocity oxy-fuel (HVOF) thermal spraying, the flame is produced via the combustion of mixed oxygen and fuel on a pressurised combustion chamber, which leaves through a nozzle creating a supersonic jet. The traditional presentation of the feedstock material is in powder form; however, that imposes a lower limit to the size of the particles to ensure adequate flowability. Suspension, and more recently, solution precursor thermal spraying have been devised as a route to avoid this limitation. In particular, solution precursor eliminates the need for suspended particles in a liquid medium. Instead, the precursors are mixed in a liquid form and then react in-flight, due to the heat transfer, to form the solid particles, consequently melting and impinging at the substrate surface (Ref 35). The technique opens up new possibilities of microstructural features and chemistries to be explored, as demonstrated by its application in the deposition of thermal barrier coatings (Ref 35,36), superhydrophobic coatings (Ref 37) or even $TiO_2$ with different porosity levels (Ref 32,33). Despite these advantages and the plethora of deposition methods reported, and to the best of the author's knowledge, no work has been published on the production of niobium-doped $TiO_2$ coatings using a solution precursor thermal spray technique.

Although a preference for titanium dioxide coatings with a high anatase content has been the norm, recent studies on the photocatalytic activity of suspension HVOF thermal sprayed titania (Ref

38) suggest that the microstructure of the deposited coating and most importantly, the interaction between rutile and anatase regions, plays an essential role in the presence of an enhanced photo-activity at relatively low anatase content (~20%). Therefore, quantitative phase content determination of XRD diffraction pattern of the deposited Nb-TiO$_2$ has been carried out in this work to better understand the formation of anatase and rutile from solution precursor feedstock.

In this work, a comprehensive study of the deposition process and microstructure of Nb-doped TiO$_2$ coatings produced using solution precursor high-velocity oxy fuel (SP-HVOF) thermal spray is presented. The process from the original precursor solution into individual droplets when exposed to the flame and the formation of the coating upon impact with the substrate, as well as the characteristics of the deposited coatings, were investigated. Scanning electron microscopy (SEM) was used to determine the morphology of the individual splats and the coatings, and x-ray diffraction was applied to evaluate relationship between the phases present, their content and the spraying parameters used. To analyse the evaporation process and high temperature behaviour of the solution precursor, thermogravimetry (TGA) and differential scanning calorimetry (DSC) analysis were performed.

## 2. Experimental methods

### 2.1. Materials and coating deposition

The solution precursor was provided by EpiValence Ltd. (Cleveland, United Kingdom) and contained a mixture of titanium ethoxide and niobium ethoxide with weight percentages of 15.0 % and 1.35 % respectively, dissolved on 2-isopropoxyethanol.

The coatings were deposited using a modified GTV TopGun HVOF system with an injector diameter of 0.3 mm directed towards a 22 mm long combustion chamber. A detailed description of the setup can be found elsewhere (Ref 39). Two sets of spraying parameters were used, corresponding to a flame power of 25 kW and 75 kW. For the 25 kW flame, the hydrogen flow rate was 78 l/min and the oxygen flow rate was 182 l/min. For the 75 kW flame, the hydrogen flow rate was 229 l/min and the oxygen flow rate was 533 l/min. In both cases the stand-off distance was 85 mm, the carousel rotation speed was 73 rpm (which corresponds to a surface speed of 1 m/s) and the gun traverse speed was 5 mm/s. 10 passes were performed to build up a coating of the desired thickness.

In order to elucidate the transformations that take place once the solution precursor enters the HVOF flame, single splats were collected on stainless steel polished substrates following a swipe test. To do so, the carousel rotation and gun traverse speed were increased to their maximum values (100 rpm and 30 mm/s respectively), while only allowing one pass of the flame. In addition, the spraying was repeated three times maintaining all spraying parameters fixed with the exception of the stand-off distance, which was chosen to be 65, 85 and 105 mm, aiming to provide three different snapshots of the evolution of the droplets as they travel along the flame.

The substrates used, with dimensions 60 x 25 x 2 mm, were AISI 304 stainless steel (SS) with nominal composition of Fe–19.0Cr–9.3Ni–0.05C (in wt. %). For the deposition of coatings, the substrates were grit blasted with a blast cleaner from Guyson (Dudley, England) with fine F100 brown alumina (0.125 - 0.149 mm) particles at 3 bar. Following grit blasting, the substrates were cleaned in industrial methylated spirit (IMS) using an ultrasonic bath for up to 10 minutes and blown dry with compressed air. In the case of single splat collection, the surface of the substrate was ground and polished down to 1 μm finish. The process was done starting with Buehler SiC grinding paper (Essligen, Germany) grit 220 (P240) until a uniform ground surface was achieved. The process was continued using grinding papers with grit 320 (P400), 400 (P800) and 600 (P1200). The polishing process was carried out using a 6 μm polishing paper, finally moving into 1 μm polishing paper for the final preparation.

2.2. <u>Characterisation</u>

Cross section of the coatings were prepared cutting a section of the substrate using a SiC cutting wheel (MetPrep Ltd, Coventry, United Kingdom) at a speed of 0.010 mm/s on an Brilliant 220 (ATM GmbH, Mammelzen, Germany) cut-off machine. The cut section was then hot-mounted using Conducto-Mount resin from MetPrep following the recommended standard procedure. The mounted cross section was then grounded and polished down to 1 μm using the same procedure as described above. Briefly, the process was done using SiC grinding papers with grits P240, P400, P800 and P1200 and 6 and 1 μm polishing papers.

A FEI Quanta 600 (FEI Europe, Eindhoven, Netherlands) scanning electron microscope (SEM) was used to image the cross section, surface and single splats of the deposited Nb-$TiO_2$, using secondary electron (SE) and backscattered electron (BSE) modes. A spot size of 2.5 nm and an acceleration voltage of 20 kV were used as the imaging parameters. The coatings were also analysed using a Siemens D500 powder X-ray diffractometer in Bragg-Brentano θ - 2θ geometry equipped with copper anode X-ray tube and a scintillation point detector. The 2θ range scanned by Cu$K\alpha$ radiation (with a wavelength of 1.54 Å) was from 20° to 120° with 0.04° step size and 22 s of counting time in each step. Peak identification was performed using the diffracsuite EVA (Bruker Software) and Rietveld refinement procedure was applied to the obtained results using TOPAS V5 software. A split pseudo-Voigt function was used to account for the base broadening of the two anatase and rutile reflections, believed to be caused by stacking faults on the crystallographic structure and some degree of amorphous content. To account for instrumental broadening effects, the specifics of the XRD instrument, such as source emission profile, detectors and slits, were defined during the refinement process. Structural values for the anatase and rutile were obtained from the inorganic Crystal Structure Database (ICSD). Since both coatings have a thickness bellow 20 µm, x-ray penetration caused the appearance of peaks from the stainless steel substrate. These peaks were taken into account during the refinement as well, although they were not considered for the total phase quantification.

In addition to Rietveld refinement, two other methods were applied to calculate the phase content of the coatings. The first one, described by Berger-Keller et al. (Ref 40) based on data from plasma-sprayed titania from powder feedstock, derives the anatase content according to the relative height of the peaks arising from the (101) reflection of the anatase phase, $I^{A(101)}$, and the (110) reflection of the rutile phase $I^{R(110)}$, following the formula shown in the Equation 1.

$$C_A = \frac{8 \times I^{A(101)}}{8 \times I^{A(101)} + 13 \times I^{R(110)}} \times 100\ \% \tag{1}$$

Due to the broadening of XRD peaks when the grain size is smaller than 300 nm (Ref 41) the use of the peak height might not be faithfully representative of the anatase content in coatings with nano-sized morphologies. To account for this phenomenon, Yang et al. (Ref 42) used the peak area of those same reflections to estimate the anatase content of suspension flame sprayed nano-$TiO_2$, using the formula shown in Equation 2.

$$C_A = \frac{A^{A(101)}}{A^{A(101)} + 1.265 \times A^{R(110)}} \times 100\ \% \tag{2}$$

Thermogravimetric analysis (TGA) and Differential scanning calorimetry (DSC) of the precursor solution were carried out in a TA Instruments SDT-Q600 thermobalance (Melbourne, Australia). A temperature profile of 10 °C/min from room temperature up to 1500 °C was applied with $Al_2O_3$ as reference and the material for the initial calibration (performed in the same conditions with an empty sample crucible). Approximately 10 mg of solution were used for the analysis, performed in an air environment. To produce dried solution for the TGA analysis, the solution precursor was kept at ~100 °C for 24 h, being the produced powder finely crushed using a mortar and pestle. The measurement parameters for the dried solution were a final temperature of 800 °C and heating rates of 10 °C/min.

3. **Results and Conclusions**

   3.1. <u>Phase evolution in SP-HVOF</u>

Two different spray runs, with flame powers 25 kW and 75 kW as described in the experimental methods section, were performed. This first set of spraying runs had the purpose of determining the effect of flame power on the microstructure and phase composition of the Nb-doped titanium oxide coatings. As it can be seen on Figure 1, the spraying with 25 kW of flame power produced a highly porous coating with loosely bonded splats. The presence of what could be identified as unpyrolysed material, not fully molten in-flight, provides the coating with numerous voids and cavities. On the other hand, the coating deposited using 75 kW flame power presents a considerably lower porosity, as shown on Figure 1, indicating that the heat transfer in-flight was enough to melt the sintered solid particles. Computational simulations of the process for 25 kW and 75 kW show that the flame at 85 mm stand-off distance reaches a temperature of around 1880 °C and 2480 °C respectively (Ref 43). These numbers support the increased degree of solid $TiO_2$ material, with a melting point of ~1840 °C, at 25 kW.

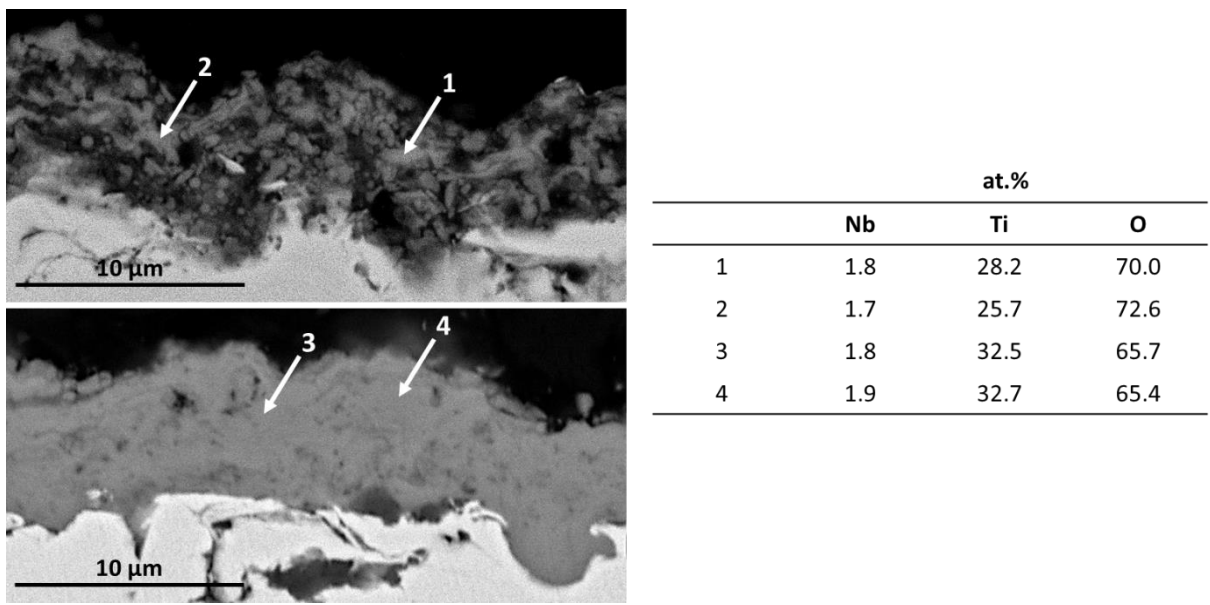

|   | at.% | | |
|---|---|---|---|
|   | Nb | Ti | O |
| 1 | 1.8 | 28.2 | 70.0 |
| 2 | 1.7 | 25.7 | 72.6 |
| 3 | 1.8 | 32.5 | 65.7 |
| 4 | 1.9 | 32.7 | 65.4 |

*Figure 1: BSE-SEM images of the cross section of the coatings at 25 kW (top) and 75 kW (bottom). The arrows point the spots where EDS measurements were taken, with the element compositions being shown in the table.*

The absence of fully molten splats during the formation of the coating at 25 kW was also confirmed through SEM imaging of the top surface of the coatings (not shown here). The surface of the 25 kW coating presents a limited amount of molten splats. Instead, smaller structures with plenty of crevices could be seen, which contribute to the high porosity. The surface of the 75 kW coating evidenced a higher percentage of particles completely molten in-flight (not shown here). This higher degree of melting allowed for a complete coverage the spaces between splats, effectively blocking the inter-splat crevices and producing a denser coating with lower porosity. This combination of molten material with solid particles can be seen on Figure 8b and Figure 8e, showing the swipe test surface of samples sprayed at 75 kW with stand-off distance of 85 mm.

### 3.1.1. Elemental distribution

In order to understand how the niobium is distributed within the coating, whether it is present as an isolated phase or as a substitutional doping element on the $TiO_2$ lattice, energy dispersive spectroscopy (EDS) measurements were taken during SEM imaging.

From the BSE images shown in Figure 1, apart from the already mentioned difference in microstructure with flame power, it can be seen that there is only a uniform distribution of Nb-doped $TiO_2$, along with porosity. The lack of brighter spots or areas, as one would expect from a heavier element such as Nb, indicates that the addition of such element causes the mentioned substitutional doping. From the EDS values it can be seen that at 25 kW, and due to the presence of unpyrolysed material, there is a larger variance for the titanium and oxygen values across the coating. The content of niobium, however, remains almost constant at 1.8 at.%. In the case of 75 kW the appearance is more uniform with no presence of unpyrolysed material. This also reflects on the titanium and oxygen content, which remains stable at different points of the cross section, matching the stoichiometry ratio of $TiO_2$. Remarkably, the niobium content also remains almost constant and equivalent to the values seen at 25 kW, which corroborates the proper doping of the $TiO_2$ with niobium regardless of the flame power and microstructure produced.

In order to analyse the effect of flame power on the phase content of the Nb-doped $TiO_2$ coatings, XRD measurements were conducted. The results, shown in Figure 2, indicate the presence of the two distinctive phases: anatase and rutile, along with some peaks corresponding to the stainless steel substrates.

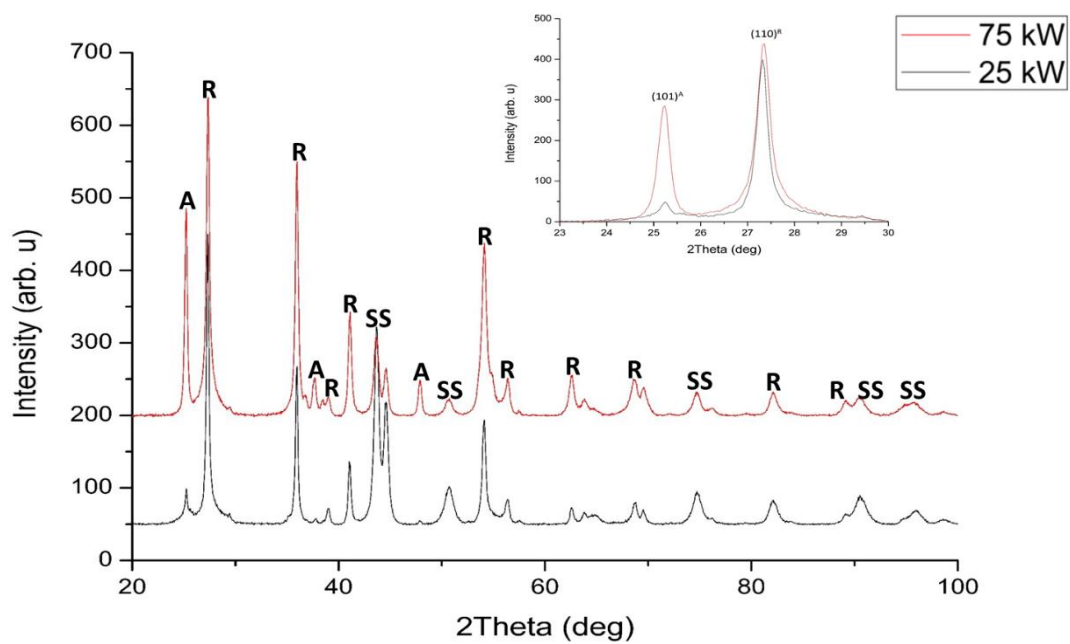

Figure 2: XRD spectra of the coatings deposited at 25 kW (red line) and 75 kW (black line). Each peak has been labelled as corresponding to Nb-TiO$_2$ rutile phase "R" (PDF card 01-072-7376), TiO$_2$

*anatase phase "A" (PDF card 00-021-1272) and 304 austenite stainless substrate "SS" (PDF card 00-033-0397). The insert shows a detailed view of the two most intense peaks for both rutile and anatase and their corresponding crystal planes.*

For the peak fitting of the XRD diffraction patterns the star quality PDF entries with reference codes 01-072-7376 (Nb-TiO$_2$, rutile phase), 00-021-1272 (TiO$_2$, anatase phase) and 00-033-0397 (304 austenite, stainless steel) were used. Unfortunately, no niobium doped anatase XRD diffraction pattern could be found in the database used (PDF-4+ 2018). The XRD spectra already indicates that the 75 kW coating presents a higher content of anatase phase.

Due to the importance of the content of both phases for the physico-chemical properties of the coating, Rietveld refinement was performed, as shown in Figure 3, to obtain the normalised phase content percentage of anatase and rutile.

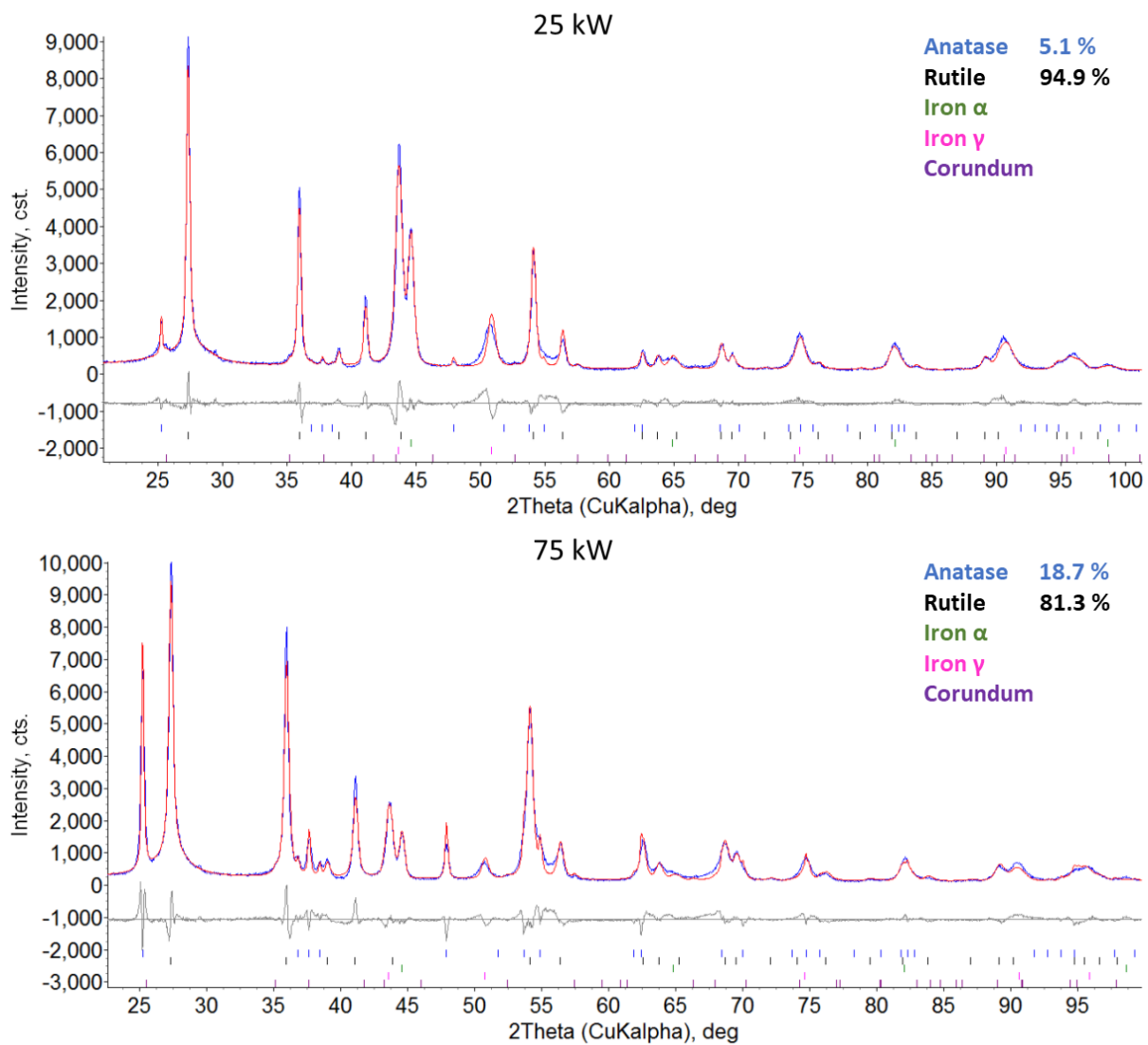

*Figure 3: Rietvield refinement of the XRD diffraction patterns with phase quantification.*

It should be noted that the Rietveld refinement process was carried out including all phases detected in the XRD diffraction patterns: the two titanium phases (anatase and rutile), those associated with the stainless steel substrate (iron α and iron γ) as well as a weak signal corresponding to corundum and associated with alumina particles from the grit blasting of the substrate. Since the thickness of the two coatings is not the same, and x-ray penetration on the sample at 25 kW was deemed to be excessive, the values presented in Figure 3 correspond to the normalised values without the substrate contribution. From the results obtained from the Rietveld refinement of both coatings it can be seen that the anatase content increases from 5.1 % to 18.7 % with a higher flame power, while the rutile content decreases from 94.9 % to 81.3 %. Some points should be mentioned regarding the phase content obtained through Rietveld refinement. The process was carried out using the structural values from standard rutile and anatase from ICSD. A slight difference is bound to occur since they correspond to undoped phases. Due to the nature of the niobium doping, entering the titanium oxide structure through substitution of titanium atoms (Ref 22), the $d_{101}$ interplanar spacing in anatase increases (Ref 28). This effect will cause a shift in the position of the anatase XRD peaks that could not be taken into account. The broadening of the first two anatase and rutile peaks, mentioned in the characterisation section, could be caused by stacking faults in the titanium oxide structure (Ref 30) and the presence of some amorphous content.

Both the height and the area of the peaks from the (101) reflection of the anatase phase and the (110) reflection of the rutile phase, shown in the inset in Figure 2, were used to calculate the anatase content of the coatings. The results of all three methods are shown summarised in Figure 4.

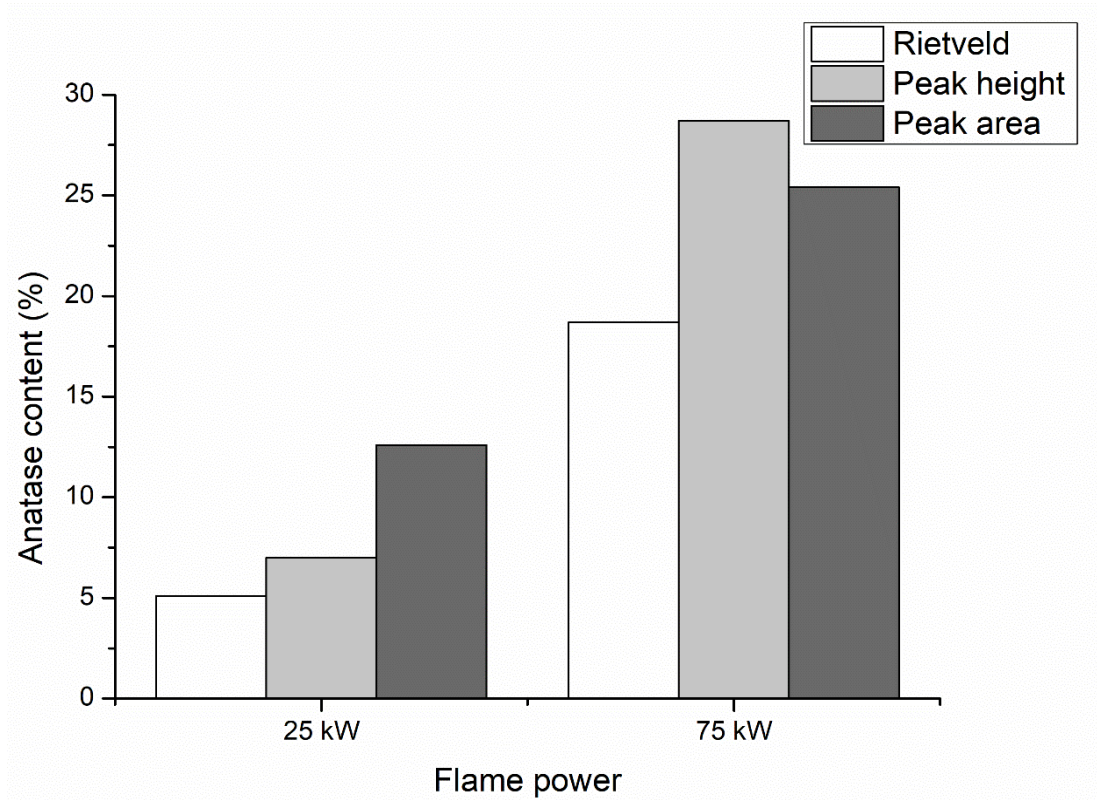

*Figure 4: Anatase content of the coatings sprayed at 25 kW (light grey) and 75 kW (dark grey) using three independent methods; Rietveld refinement, peak height and peak area.*

Although all of the three methods applied in the quantification of the anatase phase have shortcomings, the results shown in Figure 4 clearly indicate that an increase in the flame power results in an increase in the anatase content. This behaviour, contrary to the well-studied anatase-to-rutile phase transformation in titanium oxide at high temperatures (Ref 23,44,45), is intriguing. The addition of niobium to titanium oxide hinders the transition from anatase to rutile, which is explained by a reduction in oxygen vacancies when niobium enters substitutionally into $TiO_2$ (Ref 46). The reason behind this effect lies in the role of the oxygen vacancies, which act as nucleation sites for the anatase to rutile phase transformation (Ref 47). This feature, that has been exploited to maintain a more favourable phase composition for applications such as sensors (Ref 11,21), could be the reason behind the increase in the anatase phase content. However, no evidence or mechanism has been reported that estates that the addition of niobium into titanium oxide causes an alteration in the thermodynamics of the system such that at high temperature the favourable phase transition becomes rutile to anatase. Therefore, the hindering in the anatase to rutile transformation associated with niobium should be considered a reinforcing effect rather than the main reason behind this effect.

A more plausible explanation for this trend would lay in the formation mechanism of the solid content in the coating. At both 25 and 75 kW, the precursor decomposes and forms $TiO_2$ in-flight. From here, the flame temperature plays an essential role. At 25 kW, the flame temperature is approximately 1880 °C, almost identical to the melting point of TiO$_2$. As a consequence, the coating is formed by loosely bounded solid particles formed in-flight, with a mixture of rutile and anatase phases. At 75 kW, the flame temperature rises up to 2480 °C, being enough to melt the material formed in-flight. Due to the presence of air cooling at the substrates during deposition, the liquid droplets experience a high cooling rate upon impact. These conditions favour the nucleation of anatase directly from the molten material, as described by Li and Ishigaki (Ref 48), explaining the increase on anatase content.

3.2. In-flight transformation

In order to better understand the in-flight transformation of the solution precursor into solid particles, and consecutively a coating, DSC-TGA analysis was performed to gain insight on the thermodynamic behaviour of the solution. Figure 5 shows the results obtained from the analysis, where both the weight change and the heat flow from room temperature to 1500 °C are presented.

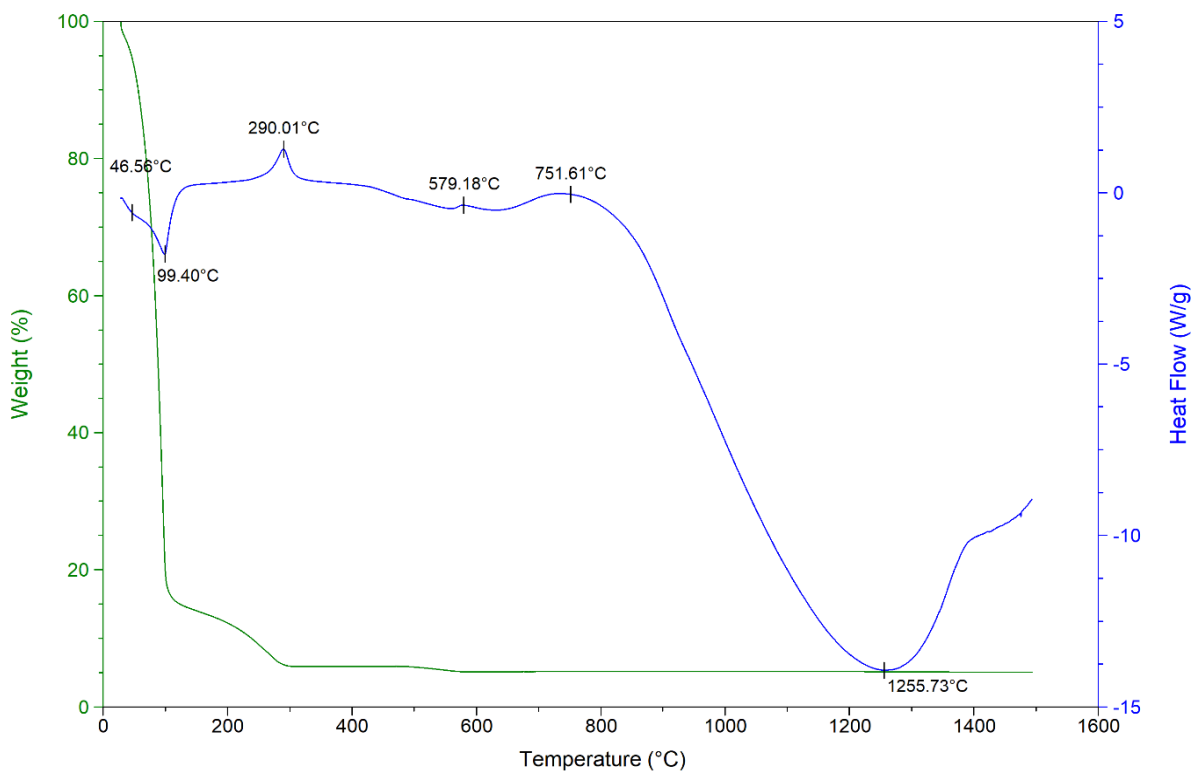

*Figure 5: TGA-DSC of the Nb-TiO$_2$ solution precursor, from room temperature to 1500 °C. The green line represents the weight change in percentage, while the blue shows the heat flow. The temperature of remarkable features in heat flow has been labelled.*

From the TGA-DSC graph it can be seen that almost 85 % of the weight is lost below 100 °C, which matches the weight proportion correspondent to the solvent. The presence of a sharp endothermic peak at 99 °C also indicates an evaporation process. Within this initial mass loss, two distinctive regions can be seen in the heat flow curve. The first one, with a shallow endothermic peak at 46.56 °C corresponds to the initial, quick evaporation of the medium. The boiling point of 2-isopropoxyethanol is 43 °C at reduced pressure (17 hPa). Due to the standard atmospheric pressure (1013 hPa) used during the TGA-DSC measurements most of the evaporation occurs at a later temperature. The second one, up to 99.40 °C also provides information about the evaporation process. Fleming *et al.* (Ref 15) performed dynamic as well as isothermal TGA measurements for the solution precursor used in this work, and found out that temperatures close to 100 °C sustained for as much as 30-35 min were needed to complete the evaporation of the solvent. This was attributed to a diffusion-limited evaporation process. A further reduction up to 95 % in weight between 100 °C and 300 °C can be seen, ending with a sharp exothermic peak at 290.01 °C. A similar peak was reported by Chen *et al.* (Ref 33) on dried precursor titanium oxide powder, being associated with the pyrolysis of the precursor. At 579.18 °C there is another exothermic peak, which matches the end of the detected mass changes by the DSC. The temperature range corresponds to previously reported phase transformation process (Ref 32).

From this point on, no distinguishable event can be seen regarding a change in weight. In the case of the heat flow, there are two additional broad peaks present, one exothermic at around 751.61 °C and a final endothermic peak at 1255.73 °C. Since no associated weight change can be detected, the origin of such peaks could be further crystallisation and phase transformation of the remaining titanium oxide dried powder.

To study in detail the pyrolysis and crystallisation of the amorphous Nb-TiO$_2$, TGA analysis of the dried solution was performed with a heating rate of 10 °C/min up to 800 °C. The results can be seen in Figure 6.

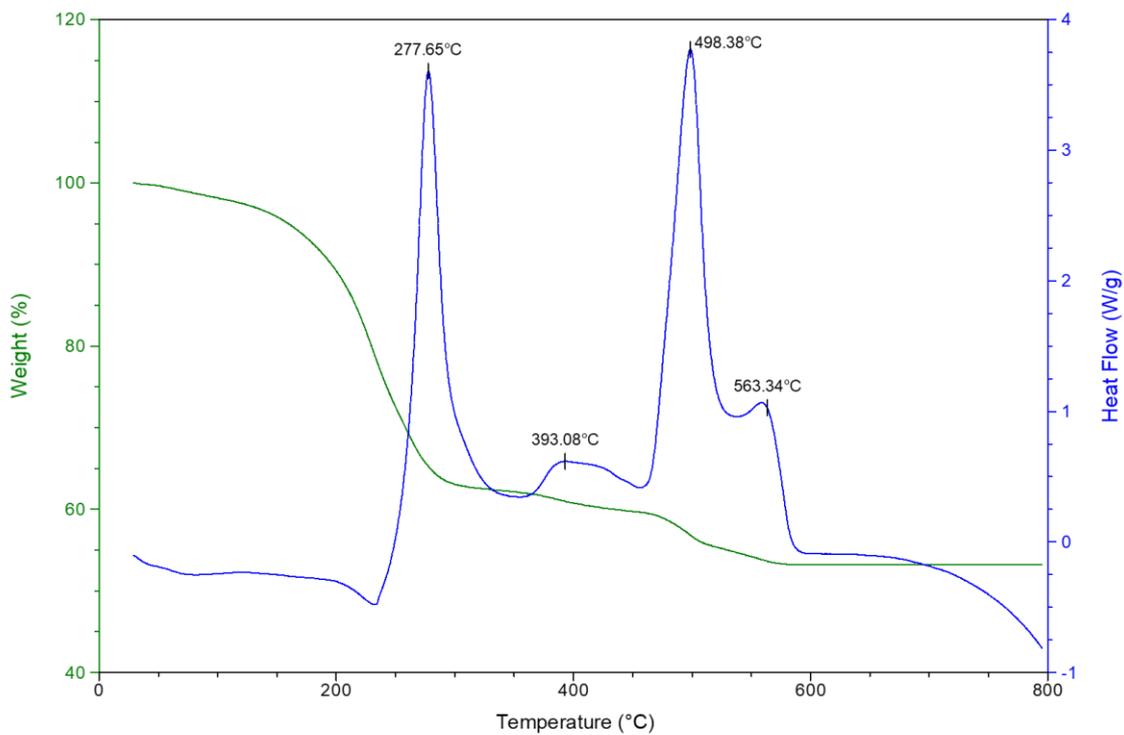

*Figure 6: TGA-DSC of the dried Nb-TiO$_2$ solution precursor, from room temperature to 800 °C.*

Since the temperature range has been reduced and the thermodynamic contribution of the solvent has been greatly removed, the data in Figure 6 provides a more clear insight into the formation of Nb-doped TiO$_2$, its crystallisation and the phase transformation. The first peak, at 277.65 °C, corresponds to the previously identified peak for the pyrolysis of the precursor. The next peak, at 393.08 °C presents a broader aspect, usually associated with a crystallisation process. Nevertheless, the complete crystallisation from amorphous phase takes place at 498.38 °C (Ref 33). The last peak, at 563.34 °C, would correspond to the transformation from anatase to rutile.

### 3.2.1. Swipe test

As mentioned in the experimental methods section, swipe tests were performed to study the morphology and presence of single splats at three distinctive stand-off distances: 65 mm, 85 mm and 105 mm. The flame power chosen was 75 kW, as it would ensure enough heat transfer to the solution in order to melt the particles formed in-flight, providing further information on the deposition mechanism. Firstly, the individual splats were analysed to understand their morphology and element composition. To do so, high magnification BSE-SEM images were analysed using EDS to understand the niobium distribution within the splats formed in-flight.

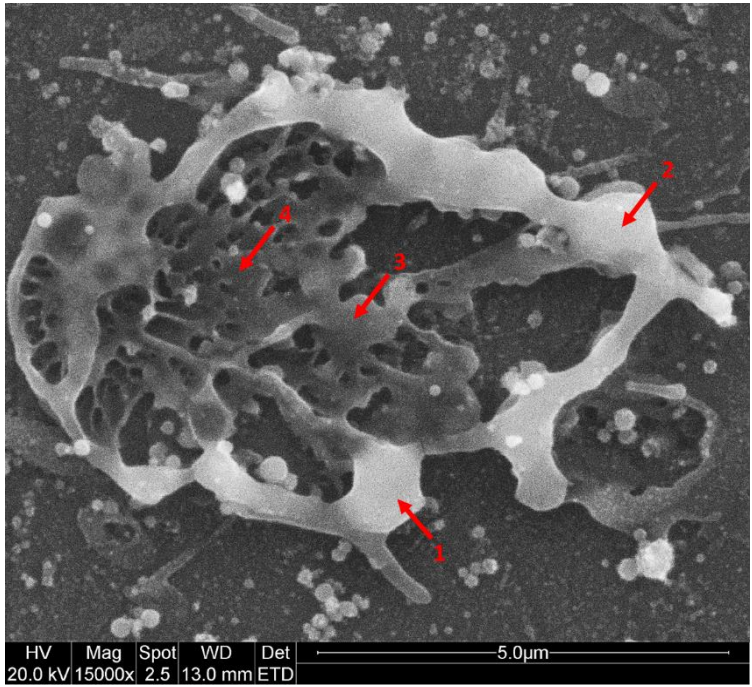

|   | at.% | | |
|---|------|---|---|
|   | Nb | Ti | O |
| 1 | 1.5 | 22.9 | 75.6 |
| 2 | 1.7 | 25.5 | 72.8 |
| 3 | 0.9 | 23.4 | 75.7 |
| 4 | 1.2 | 22.4 | 76.4 |

*Figure 7: BSE-SEM image of an individual splat. The arrows mark the points were EDS measurements were performed.*

As it can be seen in Figure 7, a liquefied splat has deformed upon impact, while preserving a brighter exterior layer and a darker interior. This difference in brightness is also reflected on the EDS measurements. Along the four different points measured, the titanium and oxygen contents remains fairly constant, being the only difference the niobium content. As expected due to the heavier nature of niobium, the brighter areas, associated with the exterior layer, present a higher content of niobium.

The top surface of the substrates, with the produced splats, is shown in Figure 8. Some features can be identified from the images, such as the reduction in deposition efficiency as the stand-off distance is increased. Nevertheless, the surface of the samples at 65 mm is mostly formed by solid particulates with a reduced presence of round, fully molten splats. Such features are more evident at 85 mm, where a mixture of solid particulates and molten splats is present. In the case of stand-off distance of 105 mm, it is considered to be excessive, as the deposition efficiency dramatically drops, without an improvement on the morphological aspect of the splats present. Some fully molten splats are still seen, accompanied by smaller, round particles.

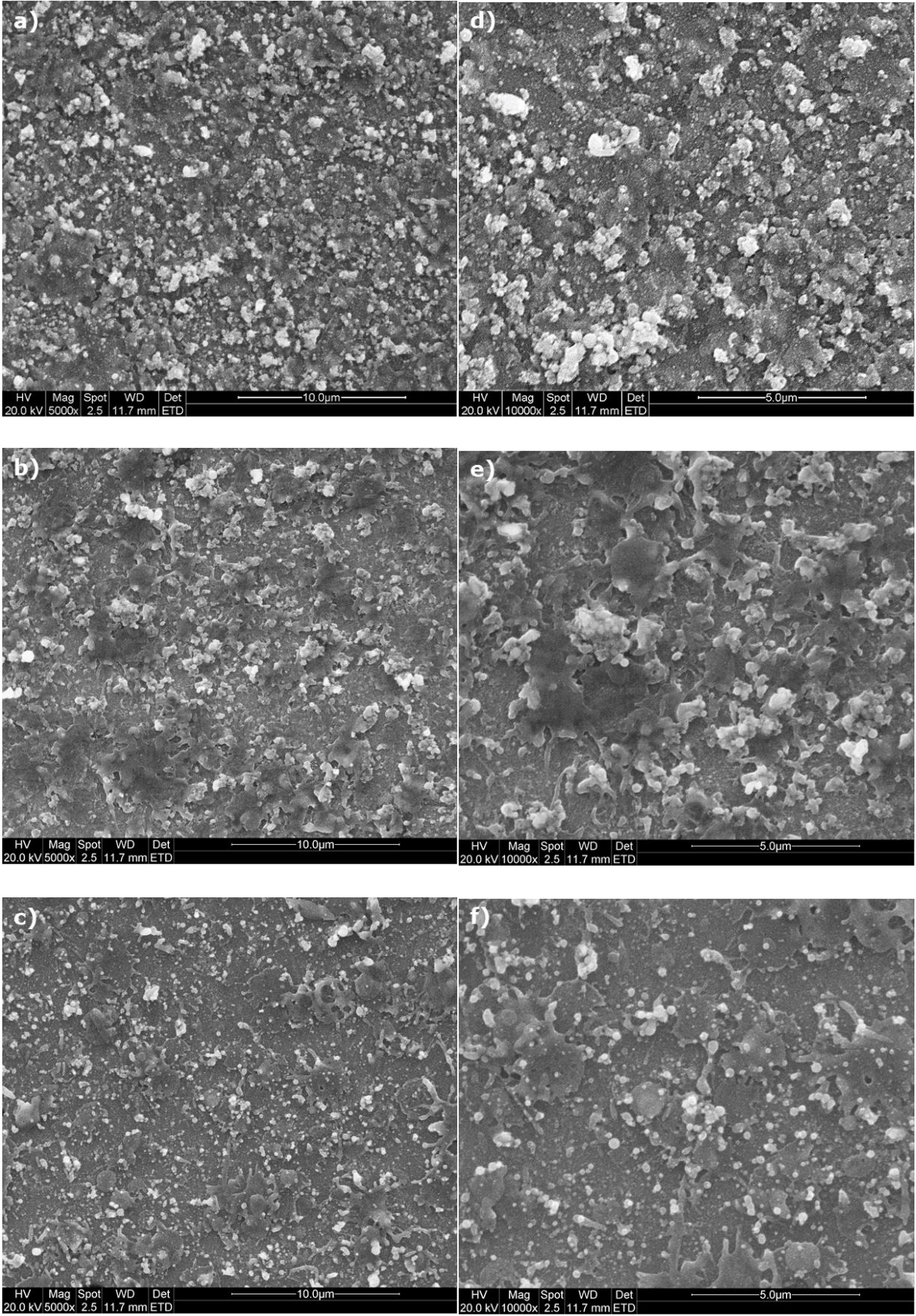

*Figure 8: SE-SEM images of the top surface of the swipe test samples. Images a to c correspond to stand-off distances of 65 mm, 85 mm and 105 mm respectively, being images d to f at high magnification. All images were taken at 20 kW.*

From the SEM images in Figure 8, it can be extracted that the amount of bright, nanosized particles is reduced as the solution precursor travels along the flame. On the other hand, the amount of molten material with relation to the solid particles increases with increased stand-off distance. This effect is partially due to the reduced deposition efficiency at larger stand-off distances; however the proportion of solid particles to molten material provides information of the physical transformation of the precursor.

A proposed model for the physical transformation that takes place once the solution precursor is injected into the system is here presented, being schematically depicted in Figure 9. The model has been developed based on the observations from the SEM images of both the cross section from coatings at different flame power and the top surface after the swipe test, and the previous reports found in the literature (Ref 49–51).

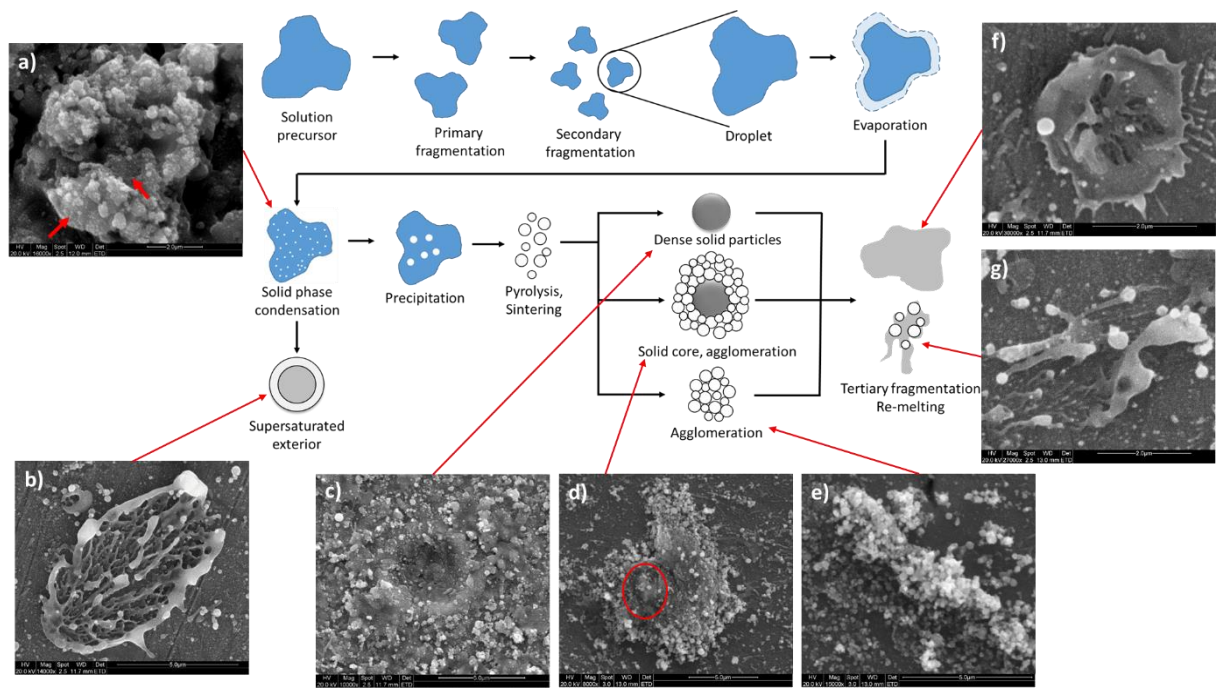

*Figure 9: Schematic diagram of the single splat formation in solution precursor HVOF of niobium doped titanium oxide. In this diagram, blue represents the solution precursor, white the Nb-rich phases, light grey the molten material and dark grey the solidified particles. BSE-SEM images correspond to the features explained in the model.*

In the schematic the process starts when the initial solution precursor is injected into the modified combustion chamber, experiencing primary fragmentation near or at the injector exit. In the case of axial injection, the presence of an annular gaseous stream surrounding the liquid jet produces a breakup into droplets due to the difference in the average velocity of the gas and the liquid (Ref 52). The specific dynamic behaviour is influenced by the dimensionless values of the Weber number (*We*) and the Reynolds number (*Re*), being defined as:

$$We = \frac{\rho v_{rel}^2 d}{\sigma}$$

$$Re = \frac{\rho v_{rel} d}{\mu}$$

Where $\rho$ is the density of the solution, $v_{rel}$ is the initial relative velocity between ambient and drop, $d$ is the initial diameter, $\sigma$ is the surface tension and $\mu$ is the dynamic viscosity. These two quantities provide a quantification of the ratio of the fluid inertia to the surface tension (*We* number) and the ratio of the fluid inertia to viscosity (*Re* number).

Once the mixture of solution precursor and gases comes out through the nozzle, entering the HVOF flame, the droplets experience a secondary fragmentation into smaller structures due to shear deformation from the drag forces (Ref 51). This process stablishes the final size of the droplets, being therefore crucial for the differentiation between the various morphologies, as illustrated in Figure 10. In this case, the *We* number is the one mostly responsible for the changes in morphology, as an increase in its value causes the droplets to experience a different secondary fragmentation from Figure 10, being the first morphology associated with lower *We* number and the last with higher values.

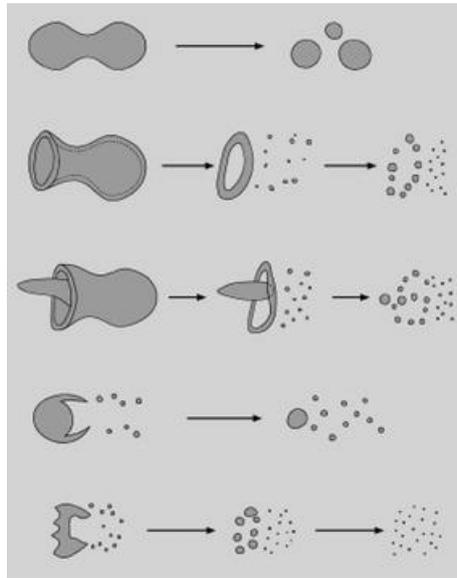

*Figure 10: Newtonian drop breakup morphologies* (Ref 53)

The interfacial tensions and viscous forces counteract the effect of the drag forces, being this equilibrium summarised in the dimensionless Ohnesorge number (*Oh*), defined as:

$$Oh = \frac{\mu}{\sqrt{\rho d \sigma}}$$

Fragmentation is favoured with an increase of the Weber number, occurring the opposite effect as the value of *Oh* increases (Ref 52).

Due to the elevated temperatures present within the flame, the solvent starts to evaporate, further reducing the size of the droplets and creating vapours within the flame. Due to the continuous heat transfer from the flame, inside the droplets starts the appearance of a condensed solid phase. This solid phase, present as bright dots within deposited splats visible only at high magnification (visible in Figure 9a, marked with arrows), was confirmed to be niobium-rich phases through energy-dispersive x-ray spectroscopy. The increased niobium content was caused by the segregation of the substitutional niobium present in the lattice of the bulk material (Ref 46,54) when exposed to high temperatures.

From the first appearance of the solid phase, two phenomena can occur. If the rate of solvent vaporisation is greater than the diffusion of the solute, a supersaturated external layer will appear surrounding the molten material, as seen in Figure 9b, and confirmed by EDS measurements on the swipe test section. In the opposite case, the condensed phases diffuse to form precipitates. This route starts with the formation of precipitates, which is followed by pyrolysis and sintering, where three morphologies can arise depending on the conditions experienced in-flight by the droplet. In the first case, if a small droplet experiences a low heat rate, a dense solid particle will form. The presence of such solid particles has been observed indirectly, as the surface of some samples from the swipe test at stand-off distance of 65 mm showed multitude of craters such as the one see in Figure 9c, indicating the impact of solid material of micrometric size. The lack of evidence for the presence of solid features on samples with larger stand-off distances indicates that these particles either lose most of its kinetic energy after traveling 65 mm, not creating craters upon impact, or they melt when traveling longer distances. Another possibility arises if the size of the initial droplets is larger or they experience higher heating rates. In this case, the formation of agglomerates of round particles will be favoured, being an example shown in Figure 9e. These agglomerated structures were more common than the craters from solid material, which provides information on the conditions of the flame and the primary and secondary fragmentation, seemingly favouring large droplets and/or higher heating rates. A third option, with characteristics in between the two previous structures was also identified and can be seen in Figure 9d. In this case, a large agglomerate possesses a solid core, marked with a circle on the BSE-SEM image. The final possible stage of the deposition process takes place when the agglomerates break up into smaller fragments, or if they never reach the critical size to endure the trajectory within the flame without re-melting, as exemplified by the images on Figure 9f and Figure 9g.

4. **Conclusions**

The SP-HVOF deposition technique was used to produce niobium-doped titanium oxide coatings for the first time. The results show that the flame power, chosen in this study to be 25 kW and 75 kW, has critical implications on the microstructural features and phase content of the produced coatings. The effects observed were:

- Flame power was proven an effective way to modify the microstructure of the deposited coatings. Low flame powers induced the sintering of solid material but failed to melt them, producing a coating with porous structure. Higher flame power allows the melting of the solid particles, leading to a lower porosity.
- Three calculation methods confirmed that an increase in flame power equated to an increase in the anatase phase content. This behaviour is believed to be caused by the presence of doped niobium, which hinders the anatase to rutile phase transformation, and the complete melting of the solid content formed in-flight at 75 kW, promoting the anatase phase content.
- In addition to the flame power, three stand-off distances (65, 85 and 105 mm) were used to collect single splats following a swipe test. The features present, as well as the morphologies discovered, were the base for the development of a model of the physico-chemical transformations that the solution precursor experiences. A detailed mechanism for the sintering of solid materials, along with the changes as it progresses through the flame, is presented.

## 5. Acknowledgments

This work was supported by the Engineering and Physical Sciences Research Council (grant number EP/L016206/1). The authors would like to thank Rory Screaton and John Kirk for their assistance during the SP-HVOF spray, Sunil Chadha for the computational simulations of the flame temperature and his input on droplet fragmentation and EpiValence Ltd. for providing the solution precursor.